\documentclass[11pt]{article}
\usepackage{amsmath} 
\usepackage[dvips]{graphics}
\usepackage{latexsym,psfig}
\usepackage{amscd}     \usepackage{amsxtra}
\usepackage{upref}     \usepackage{amsthm}
\usepackage{amssymb}   
\usepackage{amsfonts}
\begin{document}

\title{\bf Relativistic Strain and Electromagnetic Photon-Like Objects}

\author{{\bf Stoil Donev}\footnote{e-mail:sdonev@inrne.bas.bg},
{\bf Maria Tashkova}, \\
Institute for Nuclear
Research and Nuclear Energy,\\ Bulg.Acad.Sci., 1784 Sofia,
blvd.Tzarigradsko chaussee 72\\ Bulgaria\\}
\date{}
\maketitle

\begin{abstract}
This paper aims to relate some properties of photon-like objects , considered
as spatially finite time-stable physical entities with dynamical structure, to
well defined properties of the corresponding electromagnetic strains defined as
Lie derivatives of the Minkowski (pseudo)metric with respect to the eigen
vector fields of the Maxwell-Minkowski stress-energy-momentum tensor. First we
recall the geometric sense of the concept of strain, then we introduce and
discuss the notion for photon-like objects (PhLO). We compute then the strains
along the eigen vectors of the stress-energy-momentum tensor $T_{\mu}^{\nu}$
and establish important correspondences with the divergence terms of
$T_{\mu}^{\nu}$ and the terms determining some internal energy-momentum
exchange between the two recognizable component-fields $F$ and $*F$ of a vacuum
electromagnetic field. The role of appropriately defined Frobenius curvature is
also discussed and emphasized. Finally, equations of motion and interesting
PhLO-solutions are given.
 \end{abstract}


\section{Introduction}
In correspondence with  modern theoretical view on classical fields we
consider time dependent and space propagating electromagnetic fields as flows
of spatially finite physical entities which have been called in the early 20th
century {\it photons}. However, the efforts made during the past years to find
appropriate in this respect nonlinearizations of Maxwell vacuum equations
[1],[2],[3],[4],[5],[6],[7],[8],[9],[10],\newline[11],[12],[13],[14],[15],
and the seriously developed quantum theory have not resulted so far in
appropriate, from our viewpoint, description of such single time stable entities
of electromagnetic field nature with {\it finite spatial support}, in
particular, adequate theoretical image of single photons as spatially finite
physical objects with internal dynamical structure, explaining in such a way
their time stability and spin nature. As an illustration to this we give here
the Einstein's confession [16] "{\it All these fifty years of pondering have
not brought me any closer to answering the question What are light quanta}",
and this happens in front of the already developed quantim field theory. In
view of this, a decision to look back to the rudiments of the
electromagnetic theory trying to reconsider its basic assumptions in order to
come to equations giving appropriate solutions, in particular, solutions with
spatially finite carrier at every moment of their existence and
space-propagating as a whole, keeping, of course, their physical identity,
shouldn't look unjustified.

In our reconsideration and looking through the above cited attempt and review
articles and deeper studies we found some, more or less, unexplained
from theoretical viewpoint moments:

- how the {\it static} Coulomb field may be considered as acting physical object
without any self-changing: the charged mass particle changes its energy, while
the field enjoys "nontouching".

- where in Coulumb force law written as $q\mathbf{E}$ is the {\it really}
existing field generated by the charge $q$,

- should the flows of electric and magnetic fields across {\it imaginary} and
arbitrary 2-surfaces be considered as really detectable events, or the
corresponding local energy-momentum flows across appropriate physical volumes
should be used in writing corresponding balance relations,

- are the electric $\mathbf{E}$ and magnetic $\mathbf{B}$ fields are
recognizable subsystems of the time-dependent and space-propagating field, and
if such a view is accepted, how it corresponds to the fact that neither
$\mathbf{E}$ nor $\mathbf{B}$ are able to carry momentum independently of each
other,

- if the energy and momentum densities of the field are given by
$\frac12(\mathbf{E}^2+\mathbf{B}^2)$ and
$\frac1{c}\mathbf{E}\times\mathbf{B}$, then which members of the family ($f$
is a function)
$$
\Sigma_1=\mathbf{E}cos(f)+\mathbf{B}sin(f), \ \
\Sigma_2=-\mathbf{E}sin(f)+\mathbf{B}cos(f)
$$
giving the same energy-momentum densities, should be chosen for formal images
of the two subsystems,

- why the later introduced relativistic formal images $(F,*F)$ of the two
recognizable subsystems should satisfy $\mathbf{d}F=0, \mathbf{d}*F=0$, which
forbids internal energy-momentum exchange between these two subsystems in view
of the explicit divergence of the energy-momentum tensor
$$
\nabla_\nu\,T_{\mu}^{\nu}=
\nabla_\nu\Big[-\frac12\big[F_{\mu\sigma}F^{\nu\sigma} +
(*F)_{\mu\sigma}*F^{\nu\sigma}\big]\Big]=\frac12
\Big[F^{\alpha\beta}(\mathbf{d}F)_{\alpha\beta\mu}
+(*F)^{\alpha\beta}(\mathbf{d}*F)_{\alpha\beta\mu}\Big].
$$
An interesting observation based on  the formal identity
$$
\frac12F_{\mu\nu}F^{\mu\nu}\delta_{\alpha}^{\beta}=F_{\mu\alpha}F^{\mu\beta}-
(*)F_{\mu\alpha}(*F)^{\mu\beta}
$$
leading to equal energy-momentum carried by $F$ and $*F$ in the free case where
$F_{\mu\nu}F^{\mu\nu}=0$: if $F$ and $*F$ exchange locally energy-momentum
this exchange should respect the fact that the two subsystems carry equal
energy-momentum permanently.

Also, why so much interest to plane wave solutions should be paid in view of
their quite unphysical sense: infinite 3-volume supports and
corresponding infinite energy.

These and some other moments made us think that the basic equations must have
energy-momentum balance sense, and these equations should be naturally
consistent with the energy-momentum relations introduced by classical
Maxwell-Minkowski equations. During the past years we have published verious
views on the problem, the interested readers may look at the final pages of
[17]. In this paper we approach the problem from the viewpoint of elasticity:
the equations might connect local changes of the above given Minkowski
stress-energy-momentum tensor and the strain tensors on Minkowski
space-time, measuring the change of the Minkowski metric $\eta$ along
the {\it eigen vector fields} $Z_1,Z_2,...$ of $T_{\mu}^{\nu}$. In other words,
{\it outside} electromagnetic field objects the metric may be assumed to be the
flat Minkowski metric $\eta$, but {\it inside} such objects where
$T_{\mu}^{\nu}\neq 0$, a consideration of the internal local
interaction processes should take in view the values of the Lie derivatives
$L_{Z_i}(\eta), i=1,2,...$, i.e., should try to look what happens {\it inside}
in terms of such quite rapidly arising and vanishing strains/deformations,
caused by propagating electromagnetic field objects (although on unknown media,
frequently called {\it vacuum}).

\section{Strain}
The concept of {\it strain} is introduced in studying elastic materials/media
subject to external stresses of different nature: mechanical, electromagnetic,
etc. The classical strain describes mainly the abilities of the media to
bear stress-actions from outside through deformation, i.e. through changing its
shape, or, configuration.

The mathematical counterparts of the allowed deformations are those
diffeomorphisms $\varphi$ of a riemannian manifold $(M,g)$, which change the
metric $g$. It is assumed that every such $\varphi:M \rightarrow M$ generates
a possible configuration of the material considered. Since the essential
diffeomorphisms $\varphi$ must transform the metric $g$ to some new metric
$\varphi^*g$, such that $g\neq\varphi^*g$, the naturally arising tensor field
$e=(\varphi^*g-g)\neq 0$ appears as a measure of the physical abilities of the
material to withstand external force actions. It seems reasonable to recall
that these diffeomrphisms do {\it not} induce non-zero Riemannian curvature.

The continuity of the process of deformation leads to consider
1-parameter group $\varphi_t, t\in [a,b]\subset\mathbb{R}$ of deformations,
i.e. of local diffeomorphisms. Let the vector field $X$ generate $\varphi_t$,
then the quantity
$$
\frac12\,L_{X}g:=\frac12\,\lim_{t\rightarrow 0}\frac{\varphi_t^*\,g-g}{t} \ ,
$$
i.e. one half of the {\it Lie derivative} of $g$ along $X$, is called
(infinitesimal) {\it strain/deformation tensor}. \vskip
0.2cm {\bf Remark}. Further in the paper we shall work with $L_{X}\,g$,
i.e. the factor $1/2$ will be omitted. \vskip 0.2cm

Geometric approach to elasticity is given in [18], and relativistic extenstions
of the classical elasticity theory may be found in [19] and in the
corresponding quotations therein. In our further study we shall carry the above
given Lie derivative definition of strain tensor to the case of Minkowski
space-time, so we shall call $L_X\,\eta$, where $\eta$ is the Minkowski
(pseudo)metric and $X$ is any {\it eigen vector} of the Minkowski
stress-energy-momentum tensor $T_{\mu}^{\nu}$, {\it eigen strain tensor}.

Clearly, the term "material" does not seem to be appropriate
for photon-like objects (PhLO) because, as we suggest in the next section,
these objects admit NO static situations, they are of {\bf entirely dynamical
nature}, so the corresponding {\it relativistic strain} tensors must take care
of this.

\section{The Concept of photon-like object(s)}
We introduce the following notion of Photon-like object(s) (we shall use the
abbreviation "PhLO" for "Photon-like object(s)"):

\begin{center} {\bf PhLO are real massless time-stable physical objects with an
intrinsically compatible time-stable translational-rotational dynamical
structure}. \end{center}

We give now some explanatory comments, beginning with the term {\it real}. {\bf
First} we emphasize that this term means that we consider PhLO as {\it really}
existing {\it physical} objects, not as appropriate and helpful but imaginary
(theoretical) entities.  Accordingly, PhLO {\bf necessarily carry
energy-momentum}, otherwise, they could hardly be detected.  {\bf Second}, PhLO
can undoubtedly be {\it created} and {\it destroyed}, so, no point-like and
infinite models are reasonable: point-like objects are assumed to have no
structure, so they can not be destroyed since there is no available structure
to be destroyed; creation of infinite physical objects (e.g. plane waves)
requires infinite quantity of energy to be transformed from one kind to another
for finite time-period, which seems also unreasonable. Accordingly, PhLO are
{\it spatially finite} and have to be modeled like such ones, which is the only
possibility to be consistent with their "created-destroyed" nature. It seems
hardly reasonable to believe that PhLO can not be created and destroyed, and
that spatially infinite and indestructible physical objects may exist at all.
{\bf Third}, "spatially finite" implies that PhLO may carry only {\it finite
values} of physical (conservative or non-conservative) quantities.  In
particular, the most universal physical quantity seems to be the
energy-momentum, so the model must allow finite integral values of
energy-momentum to be carried by the corresponding solutions. {\bf Fourth},
"spatially finite" means also that PhLO {\it propagate}, i.e.  they do not
"move" like classical particles along trajectories, therefore, partial
differential equations should be used to describe their evolution in time.

The term "{\bf massless}" characterizes physically the way of propagation in
terms of appropriate dynamical quantities: the {\it integral}
4-momentum $p$ of a PhLO should satisfy the relation $p_\mu p^\mu=0$, meaning
that its integral energy-momentum vector {\it must be isotropic}, i.e. to
have zero module with respect to Lorentz-Minkowski (pseudo)metric in
$\mathbb{R}^4$. If the object considered has spatial and time-stable structure,
so that the translational velocity of every point where the corresponding field
functions are different from zero must be equal to $c$, we have in fact null
direction in the space-time {\it intrinsically determined} by a PhLO. Such a
direction is formally defined by a null vector field $\bar{\zeta},\bar{\zeta}^2=0$. The
integral trajectories of this vector field are isotropic (or null) {\it straight
lines} as is traditionally assumed in physics. It follows that with every PhLO a
null straight line direction is {\it necessarily} associated, so, canonical
coordinates $(x^1,x^2,x^3,x^4)=(x,y,z,\xi=ct)$ on $\mathbb{R}^4$ may be chosen
such that in the corresponding coordinate frame $\bar{\zeta}$ to have only two
non-zero components of magnitude $1$: $\bar{\zeta}^\mu=(0,0,-\varepsilon, 1)$,
where $\varepsilon=\pm 1$ accounts for the two directions along the coordinate
$z$ (further such a coordinate system will be called $\bar{\zeta}$-adapted and
will be of main usage). Our PhLO propagates {\it as a whole} along the
$\bar{\zeta}$-direction, so the corresponding energy-momentum tensor
$T_{\mu\nu}$ of the model must satisfy the corresponding {\it local isotropy
(null) condition}, namely, $T_{\mu\nu}T^{\mu\nu}=0$ (summation over the
repeated indices is throughout used).

The term "{\bf translational-rotational}" means that besides
translational component along $\bar{\zeta}$, the propagation
necessarily demonstrates some rotational (in the general sense of
this concept) component in such a way that {\it both components
exist simultaneously and consistently} with each other. It seems reasonable to
expect that such kind of behavior should be consistent only with
some distinguished spatial shapes. Moreover, if the Planck
relation $E=h\nu$ must be respected throughout the evolution, the
rotational component of propagation should have {\it
time-periodical} nature with time period $T=\nu^{-1}=h/E=const$,
and one of the two possible, {\it left} or {\it right},
orientations. It seems reasonable also to expect periodicity in
the spatial shape of PhLO, which somehow to be related to the time
periodicity.

The term "{\bf dynamical structure}" means that the propagation is supposed to
be necessarily accompanied by an {\it internal energy-momentum redistribution},
which may be considered in the model as energy-momentum exchange between (or
among) some {\it appropriately defined subsystems}.  It could also mean that
PhLO live in a dynamical harmony with the outside world, i.e.  {\it any outside
directed energy-momentum flow should be accompanied by a parallel inside
directed energy-momentum flow}.

\section{The Electromagnetic eigen strain tensors}
From formal point of view the standard relativistic formulation of classical
electrodynamics in vacuum (zero charge density) is based on the following
assumptions. The configuration space is the Minkowski space-time
$M=(\mathbb{R}^4,\eta)$ where $\eta$ is the pseudometric with
$sign(\eta)=(-,-,-,+)$ with the corresponding volume 4-form $\omega_o=dx\wedge
dy\wedge dz\wedge d\xi $ and Hodge star $*$ defined by
$\alpha\wedge *\beta=(-1)^{ind\,\eta}\eta(\alpha,\beta)\omega_o$,
$\alpha$ and $\beta$ are forms of the same degree. The electromagnetic field is
described by two closed 2-forms $(F,*F):\mathbf{d}F=0, \ \mathbf{d}*F=0$. The
physical characteristics of the field are deduced from the following
stress-energy-momentum tensor field
\begin{equation}
T_{\mu}{^\nu}(F,*F)=-\frac12\big[F_{\mu\sigma}F^{\nu\sigma}+
(*F)_{\mu\sigma}(*F^{\nu\sigma})\big].
\end{equation}
Note that there is NO interaction stress-energy term between $F$ and $*F$ .
In the non-vacuum case the allowed energy-momentum exchange with other
physical systems is given in general by the divergence
\begin{equation}
\nabla_\nu\,T_{\mu}^{\nu}
=\frac12 \Big[F^{\alpha\beta}(\mathbf{d}F)_{\alpha\beta\mu}
+(*F)^{\alpha\beta}(\mathbf{d}*F)_{\alpha\beta\mu}\Big] =
F_{\mu\nu}(\delta F)^\nu + (*F)_{\mu\nu}(\delta *F)^\nu,
\end{equation}
where $\delta=*\mathbf{d}*$ is the co-derivative. Since
$\mathbf{d}F=0, \mathbf{d}*F=0$, this divergence is obviously
equal to zero on the vacuum solutions: its both terms are zero.
Therefore, energy-momentum exchange between the two
subsystem-fields $F$ and $*F$, as suggested by the above divergence terms,
should be expressed by
$$
(*F)^{\alpha\beta}(\mathbf{d}F)_{\alpha\beta\mu}, \ \ \
F^{\alpha\beta}(\mathbf{d}*F)_{\alpha\beta\mu},
$$
summation over $\alpha<\beta$, is NOT allowed on the
solutions of $\mathbf{d}F=0, \mathbf{d}*F=0$. This shows that the
widely used 4-potential approach (even if two 4-potential 1-forms $U,U^*$ are
introduced so that $\mathbf{d}U=F, \ \mathbf{d}U^*=*F$ locally)
to these equations forbids real (coordinate free) changes, and excludes any
possibility to individualize two energy-momentum exchanging time-recognizable
subsystems of the field that are mathematically represented by $F$ and $*F$.

According to the above introduced and discussed concept of PhLO we
have to impose the condition $T_{\mu\nu}T^{\mu\nu}=0$ on the
energy tensor (1). As is well known [20] this is equivalent to
zero eigen values of $T_{\mu}^{\nu}$, which implies zero
invariants: $I_{1}=\frac12F_{\mu\nu}F^{\mu\nu}=
\frac12F_{\mu\nu}(*F)^{\mu\nu}=0$, and that $T_{\mu}^{\nu}$ admits
just one null eigen direction defined by the vector field
$\bar{\zeta}$, $\bar{\zeta}^2=0$, determining a null straight-line
direction along which the energy density propagates
translationally. This isotropic nature of the field leads [20] to the
following structure of $F$ and $*F$
$$
F=A\wedge\zeta, \ *F=A^*\wedge\zeta,
$$
where $\zeta_{\mu}=\eta_{\mu\nu}\bar{\zeta}^{\nu}$,  $A$ and $A^*$ are the
spatial-like electric and magnetic 1-form components of the field: $A^2<0,
(A^*)^2<0$, $\eta(A,A^*)=0$, moreover, $(\bar{A},\bar{A^*})$ are eigen
vectorsof $T_\mu^\nu$, and $\eta(A,\zeta)=\eta(A^*,\zeta)=0$. Under these
conditions it is possible to find, and convenient to use, a canonical
coordinate system $(x,y,z,\xi=ct)$ such that $\zeta =\varepsilon dz +d\xi$,
$A=u\,dx+p\,dy $, $A^*=\varepsilon\,p\,dx-\varepsilon u\,dy$, where $(u,p)$ are
two functions, and $\varepsilon=\pm 1$ carries information about the direction
of propagation along the coordinate $z$ of our PhLO.

{\bf Remark} All further identifications of tangent and
cotangent objects will be made by $\eta$, and if $A$ is a tangent coobject
 we shall denote by $\bar{A}$ the $\eta$-corresponding tangent object.
\vskip 0.3cm
Since the three vector fields
$(\bar{\zeta},\bar{A},\bar{A}^*)$ are eigen vectors of $T_{\mu}^{\nu}$, we
compute the corresponding three eigen strain tensors.

{\bf Proposition}. The following relations hold:
 \[ \begin{split}
L_{\bar{\zeta}}\,\eta &=0, \qquad
(L_{\bar{A}}\,\eta)_{\mu\nu}\equiv
D_{\mu\nu}= \begin{Vmatrix}2u_x & u_y+p_x & u_z & u_{\xi} \\
u_y+p_x & 2p_y & p_z & p_{\xi} \\ u_z & p_z & 0 & 0 \\ u_{\xi} &
p_{\xi} & 0 & 0  \end{Vmatrix} , \\
(L_{\bar{A^*}}\,\eta)_{\mu\nu} &\equiv D^*_{\mu\nu} =
\begin{Vmatrix}-2\varepsilon p_x & -\varepsilon(p_y+u_x) & -\varepsilon p_z &
-\varepsilon p_{\xi} \\ -\varepsilon(p_y+u_x) & 2\varepsilon u_y & \varepsilon
u_z & \varepsilon u_{\xi} \\ -\varepsilon p_z & \varepsilon u_z & 0 & 0 \\
-\varepsilon p_{\xi} & \varepsilon u_{\xi} & 0 & 0 \end{Vmatrix} .
\end{split}
\]

{\bf Proof}. Immediately verified.
\vskip 0.5cm

Let $[X,Y]$ denote the Lie bracket of vector fields on $M$, and
$\phi^2=u^2+p^2$, $\psi=\mathrm{arctg}(p/u)$, i.e.
$u=\phi\,\mathrm{cos}\psi, \ p=\phi\,\mathrm{sin}\psi$. We note
that the local conservation law $\nabla_{\nu}T_{\mu}^{\nu}=0$
reduces in our case to $L_{\bar{\zeta}}\phi^2=0$, so $(u,p)$ would
be running waves if $L_{\bar{\zeta}}\psi=0$ too.

We give now some important from our viewpoint relations.
\begin{eqnarray}
D(\bar{\zeta},\bar{\zeta})& =& D^*(\bar{\zeta},\bar{\zeta})=0,
\nonumber \\
D(\bar{\zeta}) &\equiv & D(\bar{\zeta})_\mu dx^\mu\equiv
D_{\mu\nu}\bar{\zeta}^\nu dx^\mu =(u_\xi-\varepsilon u_z)dx +
(p_\xi-\varepsilon p_z)dy, \nonumber
\\
D(\bar{\zeta})^\mu\frac{\partial}{\partial x^\mu} &\equiv &
D^{\mu}_{\nu}\bar{\zeta}^\nu\frac{\partial}{\partial x^\mu}=
-(u_\xi-\varepsilon u_z)\frac{\partial}{\partial x} -
(p_\xi-\varepsilon p_z)\frac{\partial}{\partial
y}=-[\bar{A},\bar{\zeta}], \nonumber
\\
D_{\mu\nu}\bar{A}^\mu\bar{\zeta}^\nu &=&
-\frac12\Big[(u^2+p^2)_\xi -\varepsilon(u^2+p^2)_z\Big]=
-\frac12L_{\bar{\zeta}}\phi^2 , \\
D_{\mu\nu}\bar{A^*}^\mu\bar{\zeta}^\nu &=&
-\varepsilon\Big[u(p_\xi-\varepsilon p_z)-p(u_\xi-\varepsilon
u_z)\Big]= -\varepsilon\mathbf{R}=-\varepsilon
\phi^2\,L_{\bar{\zeta}}\psi,
\end{eqnarray}
where $\mathbf{R}\equiv u(p_\xi-\varepsilon
p_z)-p(u_\xi-\varepsilon u_z)$. We also have:
\begin{eqnarray}
D^*(\bar{\zeta}) &=&\varepsilon\Big[-(p_\xi-\varepsilon p_z)dx+
(u_\xi-\varepsilon u_z)dy\Big] , \nonumber
\\
D^*(\bar{\zeta})^\mu\frac{\partial}{\partial x^\mu} &\equiv &
(D^*)^{\mu}_{\nu}\bar{\zeta}^\nu\frac{\partial}{\partial x^\mu}=
-\varepsilon(p_\xi-\varepsilon p_z)\frac{\partial}{\partial x} +
(u_\xi-\varepsilon u_z)\frac{\partial}{\partial
y}=[\bar{A^*},\bar{\zeta}], \nonumber
\\
D^*_{\mu\nu}\bar{A^*}^\mu\bar{\zeta}^\nu &=&
-\frac12\Big[(u^2+p^2)_\xi -\varepsilon(u^2+p^2)_z\Big]=
-\frac12L_{\bar{\zeta}}\phi^2 ,\\
D^*_{\mu\nu}\bar{A}^\mu\bar{\zeta}^\nu &=&
\varepsilon\Big[u(p_\xi-\varepsilon p_z)-p(u_\xi-\varepsilon
u_z)\Big]= \varepsilon\mathbf{R}=\varepsilon
\phi^2\,L_{\bar{\zeta}}\psi.
\end{eqnarray}
Clearly, $D(\bar{\zeta})$ and $D^*(\bar{\zeta})$ are linearly
independent in general:
$$
D(\bar{\zeta})\wedge D^*(\bar{\zeta})=\varepsilon
\Big[(u_\xi-\varepsilon u_z)^2+(p_\xi-\varepsilon p_z)^2\Big]dx\wedge dy
=\varepsilon\phi^2(\psi_\xi-\varepsilon \psi_z)^2\,dx\wedge dy\neq 0.
$$

We readily obtain now
\begin{multline*} D(\bar{\zeta})\wedge F=D^*(\bar{\zeta})\wedge
(*F)= D(\bar{\zeta})\wedge A\wedge\zeta=D^*(\bar{\zeta})\wedge
A^*\wedge\zeta \\
=-\varepsilon\Big[u(p_\xi-\varepsilon p_z)- p(u_\xi-\varepsilon
u_z)\Big]dx\wedge dy\wedge dz \\ - \Big[u(p_\xi-\varepsilon p_z)-
p(u_\xi-\varepsilon u_z)\Big]dx\wedge dy\wedge d\xi
\\
=-\phi^2\,L_{\bar{\zeta}}\psi\,(\varepsilon\,dx\wedge dy\wedge dz+
dx\wedge dy\wedge d\xi) =-\mathbf{R}\,(\varepsilon\,dx\wedge
dy\wedge dz+ dx\wedge dy\wedge d\xi), \end{multline*}
\[ \begin{split} D(\bar{\zeta})\wedge
(*F)&=-D^*(\bar{\zeta})\wedge F= D(\bar{\zeta})\wedge
A^*\wedge\zeta=-D^*(\bar{\zeta})\wedge A\wedge\zeta
\\
&=\frac12\Big[(u^2+p^2)_\xi-\varepsilon(u^2+p^2)_z\Big] (dx\wedge
dy\wedge dz+\varepsilon\,dx\wedge dy\wedge d\xi). \end{split} \]
Thus we get
\begin{equation}
*\Big[D(\bar{\zeta})\wedge
A\wedge\zeta\Big]= *\Big[D^*(\bar{\zeta})\wedge A^*\wedge\zeta\Big]=
-\varepsilon\mathbf{R}\,\zeta =-i(\bar{*F})\mathbf{d}F=i(\bar{F})\mathbf{d}(*F),
\end{equation}
\begin{equation}
*\Big[D(\bar{\zeta})\wedge
A^*\wedge\zeta\Big]= -*\Big[D^*(\bar{\zeta})\wedge A\wedge\zeta\Big]=
\frac12L_{\bar{\zeta}}\phi^2\,\zeta=
i(\bar{F})\mathbf{d}F=i(\bar{*F})\mathbf{d}(*F),
\end{equation}
where, if $(K,G)$ are respectively a 2-form and a 3-form, we have denoted
$$
i(\bar{K})G=K^{\mu\nu}G_{\mu\nu\sigma}dx^{\sigma}, \ \mu<\nu .
$$

\section{Discussion}
The above relations show various dynamical aspects of the energy-momentum
redistribution during evolution of our PhLO in terms of the
{\it intrinsically} defined electromagnetic strains.

Relation (8) shows how the local conservation law
$\nabla_{\nu}T_{\mu}^{\nu}=0$, being equivalent to
$L_{\bar{\zeta}}\phi^2=0$, and meaning that the energy density
$\phi^2$ propagates just translationally along $\bar{\zeta}$, can
be expressed in terms of the eigen strain components. On the other hand
the phase function $\psi$ would satisfy $L_{\bar{\zeta}}\psi=0$
ONLY if the introduced quantity $\mathbf{R}$ is equal to zero:
$\mathbf{R}=0$. Now, what is the sense of $\mathbf{R}$ reveals the
following

{\bf Proposition}. The following relations hold:
\[ \begin{split}
\mathbf{d}A\wedge A\wedge A^* &=0;\qquad
\mathbf{d}A^*\wedge A^*\wedge A=0;                        
\\ \mathbf{d}A\wedge A\wedge \zeta &=
\varepsilon\big[u(p_\xi-\varepsilon p_z)- p(u_\xi-\varepsilon
u_z)\big]\omega_o; \\
\mathbf{d}A^*\wedge A^*\wedge \zeta &=                                  
\varepsilon\big[u(p_\xi-\varepsilon p_z)- p(u_\xi-\varepsilon
u_z)\big]\omega_o. \end{split}
\]
{\bf Proof}. Immediately verified.

These relations say that the 2-dimensional Pfaff system $(A,A^*)$
is completely integrable for any choice of the two functions
$(u,p)$, while the two 2-dimensional Pfaff systems $(A,\zeta)$ and
$(A^*,\zeta)$ are NOT completely integrable in general, and the
same curvature factor $$ \mathbf{R}=u(p_\xi-\varepsilon
p_z)-p(u_\xi-\varepsilon u_z) $$ determines their
nonintegrability. Hence, the condition $\mathbf{R}\neq 0$ allows
in principle rotational component of propagation, i.e. $u$ and $p$
NOT to be plane waves, so compatibility between the translational
and rotational components of propagation is possible in principle.

As we mentioned above it is possible the translational and rotational
components of the energy-momentum redistribution to be represented in form
depending on the $\bar{\zeta}$-directed strains $D(\bar{\zeta})$ and
$D^*(\bar{\zeta})$. So, the local translational changes of the energy-momentum
carried by the two components $F$ and $*F$ of our PhLO are given by the two
1-forms $*\big[D(\bar{\zeta})\wedge A^*\wedge\zeta\big]$ and
$*\big[D^*(\bar{\zeta})\wedge A\wedge\zeta\big])$, and the local rotational ones
are given by the 1-forms $*\big[D(\bar{\zeta})\wedge A\wedge\zeta\big]$ and
$*\big[D^*(\bar{\zeta})\wedge A^*\wedge\zeta\big]$.  In fact, the 1-form
$*\big[D(\bar{\zeta})\wedge A\wedge\zeta\big]$ determines the strain that tends
to "leave" the 2-plane defined by $(A,\zeta)$ and the 1-form
$*\big[D^*(\bar{\zeta})\wedge A^*\wedge\zeta\big]$ determines the strain that
tends to "leave" the 2-plane defined by $(A^*,\zeta)$. So, the
particular kind of nonintegrability of $(A,\zeta)$ and $(A^*,\zeta)$
mathematically guarantees some physical inter-communication through
establishing local dynamical equilibrium (eq.(7))
between two subsystems of the
field, that are mathematically represented by $F$ and $*F$.

Since the PhLO is free,
i.e. no energy-momentum is lost or gained, this means that the two (null-field)
components $F$ and $*F$ exchange locally {\it equal} energy-momentum
quantities:
$$
*\Big[D(\bar{\zeta})\wedge
A\wedge\zeta\Big]= *\Big[D^*(\bar{\zeta})\wedge A^*\wedge\zeta\Big] .
$$
Now, the local energy-momentum conservation law
\[ \nabla_{\nu}\big[F_{\mu\sigma}F^{\nu\sigma}+
(*F)_{\mu\sigma}(*F)^{\nu\sigma}\big]=0,\] i.e.
$L_{\bar{\zeta}}\phi^2=0$, requires the corresponding eigen strain-fluxes
 $*\big[D^*(\bar{\zeta})\wedge A\wedge\zeta\big]$ and
$*\big[D(\bar{\zeta})\wedge A^*\wedge\zeta\big]$ to become zero:
$$
*\Big[D(\bar{\zeta})\wedge
A^*\wedge\zeta\Big]= -*\Big[D^*(\bar{\zeta})\wedge A\wedge\zeta\Big]=
\frac12L_{\bar{\zeta}}\phi^2\,\zeta=
i(\bar{F})\mathbf{d}F=i(\bar{*F})\mathbf{d}(*F)=0 ,
$$
and the local dynamical balance between $F$ and $*F$ to hold:
$$
*\Big[D(\bar{\zeta})\wedge
A\wedge\zeta\Big]= *\Big[D^*(\bar{\zeta})\wedge A^*\wedge\zeta\Big]=
-\varepsilon\mathbf{R}\,\zeta
=-i(\bar{*F})\mathbf{d}F=i(\bar{F})\mathbf{d}(*F).
$$

It seems important to note that, only differential relation between
the energy-momentum change and natural strain fluxes exists, so NO analog
of the assumed in elasticity theory generalized Hooke law, i.e.
linear relation between the stress tensor and the strain tensor,
seems to exist here. This clearly goes along with the
fully dynamical nature of PhLO, i.e. linear relations exist
between the divergence terms of our stress tensor
$\frac12\big[-F_{\mu\sigma}F^{\nu\sigma}-
(*F)_{\mu\sigma}(*F)^{\nu\sigma}\big]$ and the correspondingly
directed eigen strain fluxes as given by equations (2,7,8).

The constancy of the translational propagation and the required
translational-rotational compatibility suggest also constancy of of
the rotational component of propagation, i.e.
$L_{\bar{\zeta}}\psi= const$. So, our equations
$\Big[D(\bar{\zeta})\wedge A^*\wedge\zeta\Big]=
-\Big[D^*(\bar{\zeta})\wedge A\wedge\zeta\Big]=0, \
\Big[D(\bar{\zeta})\wedge A\wedge\zeta\Big]=
\Big[D^*(\bar{\zeta})\wedge A^*\wedge\zeta\Big]$, or in terms of $F$:
$i(\bar{F})\mathbf{d}F=i(\bar{*F})\mathbf{d}(*F)=0, \ \
i(\bar{*F})\mathbf{d}F+i(\bar{F})\mathbf{d}(*F)=0$ reduce to
$$
L_{\bar{\zeta}}\phi^2=0, \qquad L_{\bar{\zeta}}\psi=
\frac{\kappa}{l_o},
$$
where $\kappa=\pm1$ and $l_o$ is a parameter of dimension of length. Among
the nonlinear solutions of these equations we find out the following:
\[
\begin{split}
u(x,y,z,\xi)
&=\phi(x,y,\xi+\varepsilon z)\,
\mathrm{cos}(-\varepsilon\kappa\frac{z}{l_o}+ const), \\ p(x,y,z,\xi)
&=\phi(x,y,\xi+\varepsilon z)\, \mathrm{sin}(-\varepsilon\kappa\frac{z}{l_o}+
const), \end{split}
\]
where $\phi(x,y,\xi+\varepsilon z)$ is an arbitrary smooth function, so it can
be chosen to have 3d-finite spatial support inside an appropriate 3-region
such, that at every moment our PhLO to be localized inside an one-step long
helical cylinder wrapped up around the $z$-axis.

On the two figures below are given two theoretical examples with $\kappa=-1$
and $\kappa=1$ respectively, at a fixed moment $t$. For $t\in
(-\infty,+\infty)$, the amplitude function $\phi$ fills in a smoothed out tube
around a circular helix of height $2\pi\mathcal{L}_o$ and pitch
$\mathcal{L}_o$, and phase function $\varphi=\mathrm{cos}(\kappa
z/\mathcal{L}_o)$. The solutions propagate left-to-right along the euclidean
coordinate $z$.

\vskip0.5cm
\begin{center} \begin{figure}[ht!] \centerline{
{\mbox{\psfig{figure=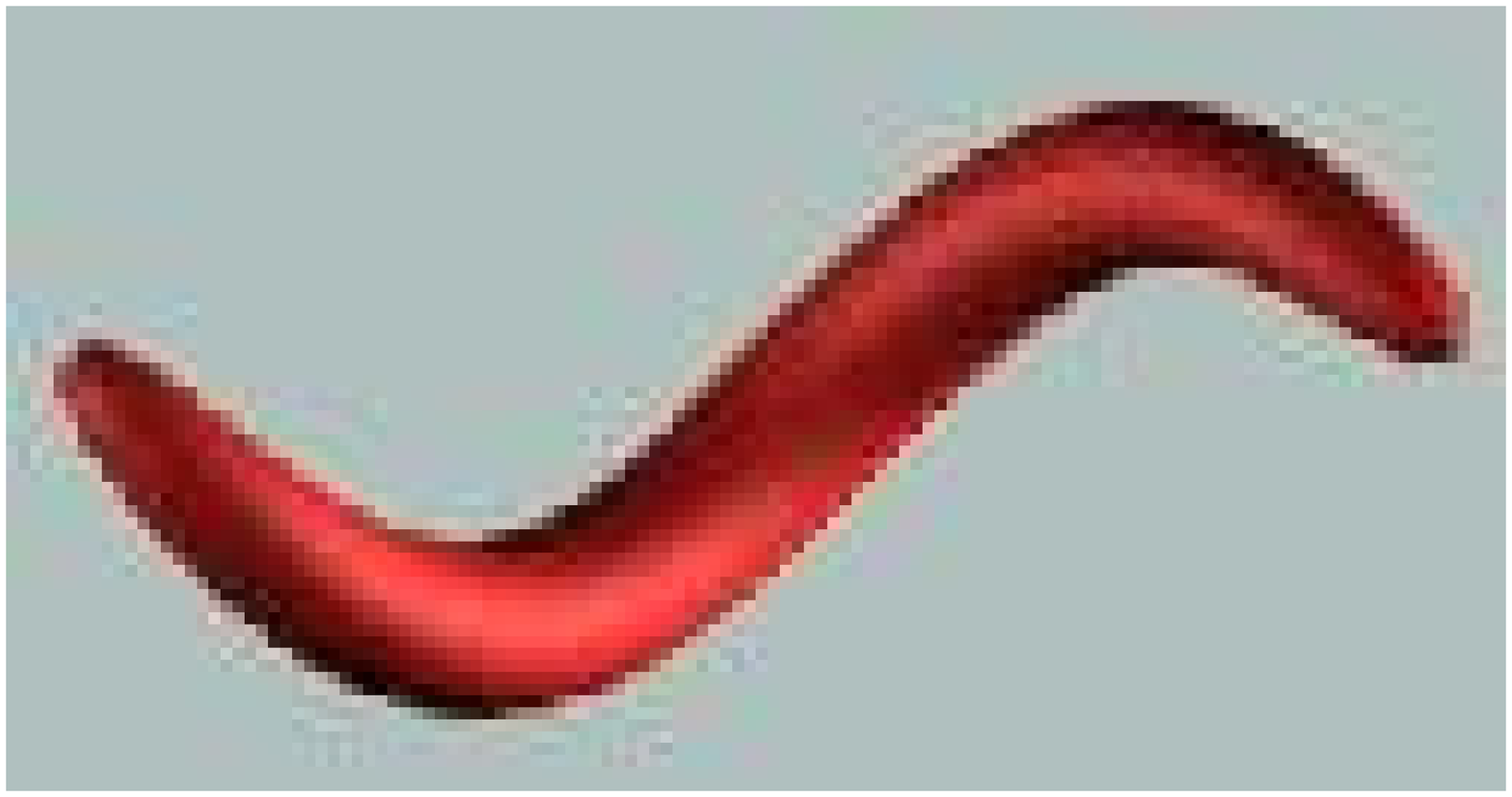,height=1.5cm,width=3.5cm}}
\mbox{\psfig{figure=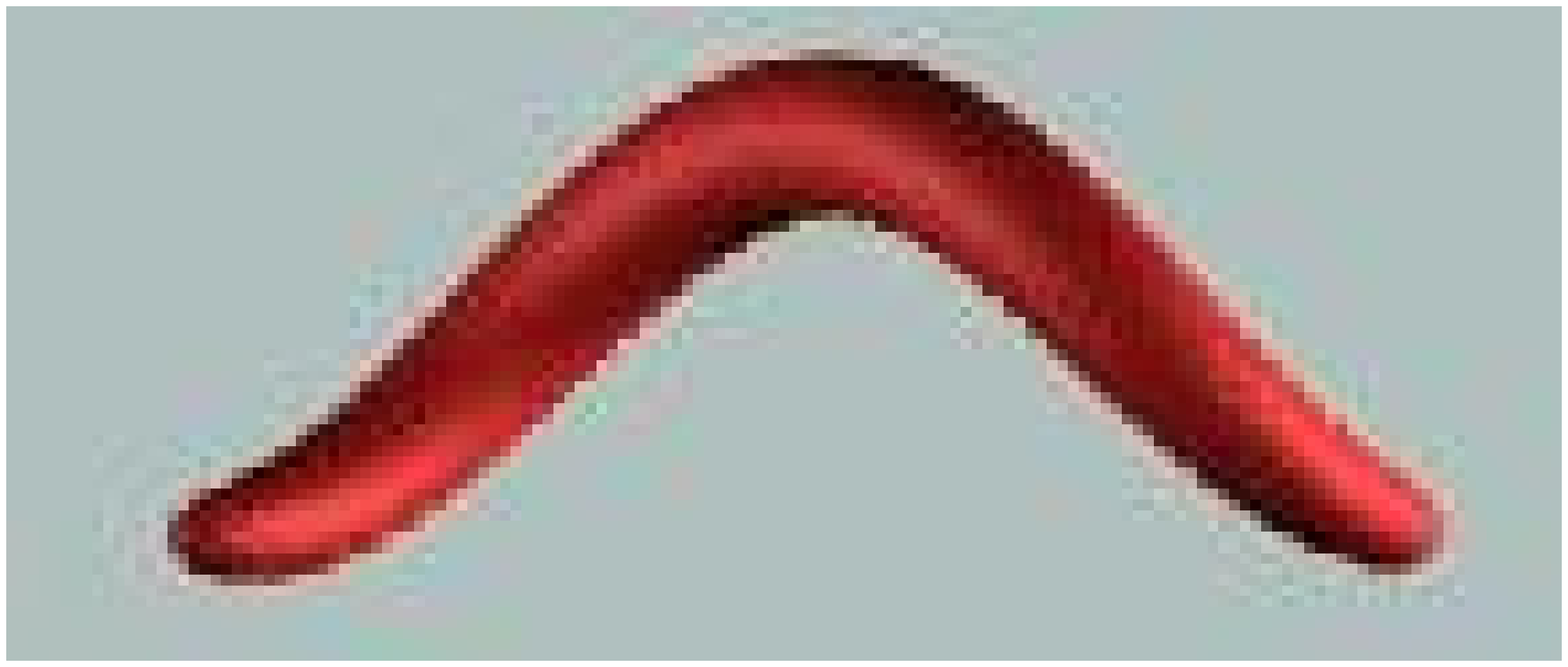,height=1.5cm,width=4.2cm}}
\mbox{\psfig{figure=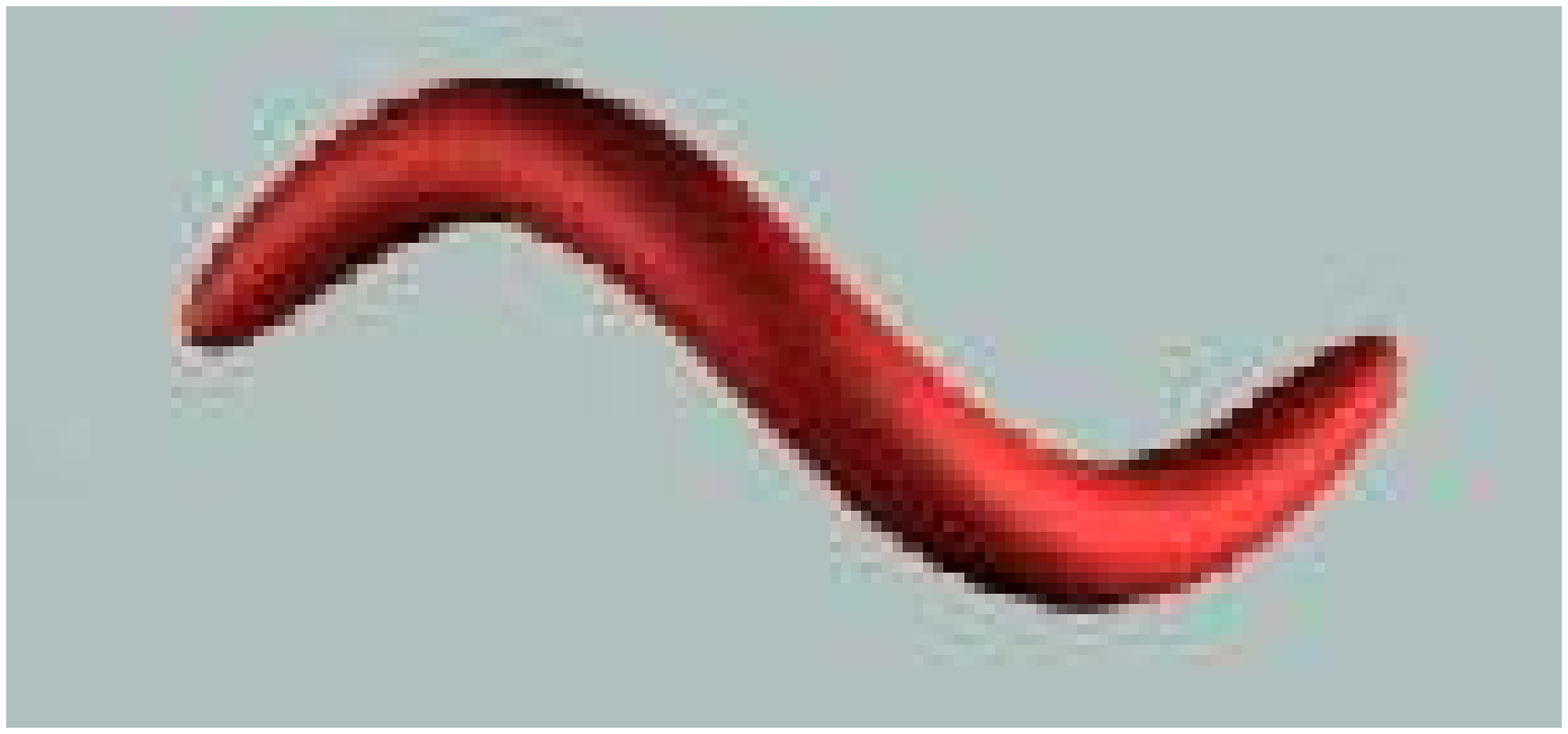,height=1.5cm,width=4.2cm}}}}
\caption{Theoretical example with $\kappa=-1$. The translational
propagation is directed left-to-right.}
\end{figure}
\end{center}
\begin{center}
\begin{figure}[ht!]
\centerline{
{\mbox{\psfig{figure=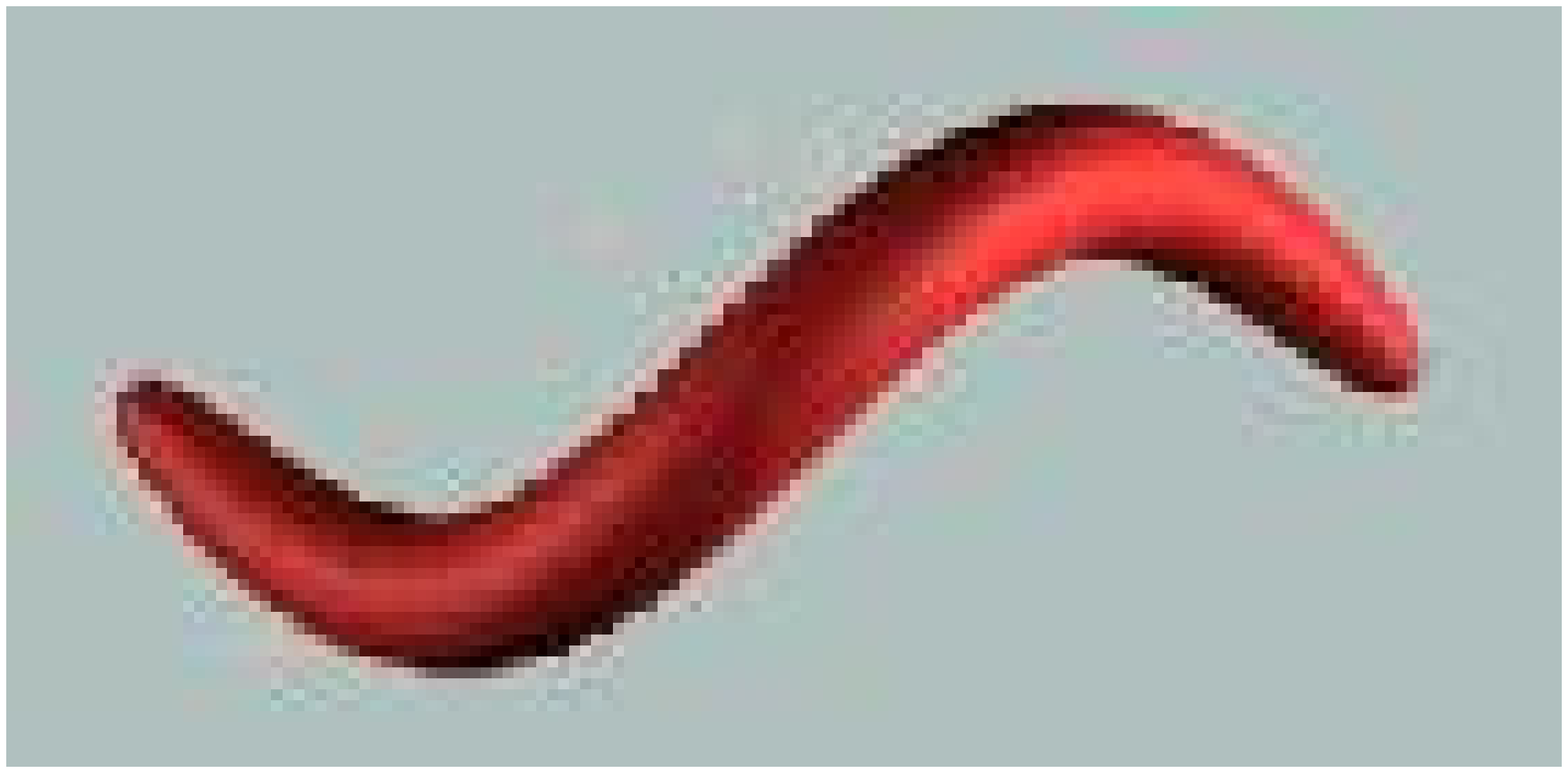,height=1.5cm,width=3.5cm}}
\mbox{\psfig{figure=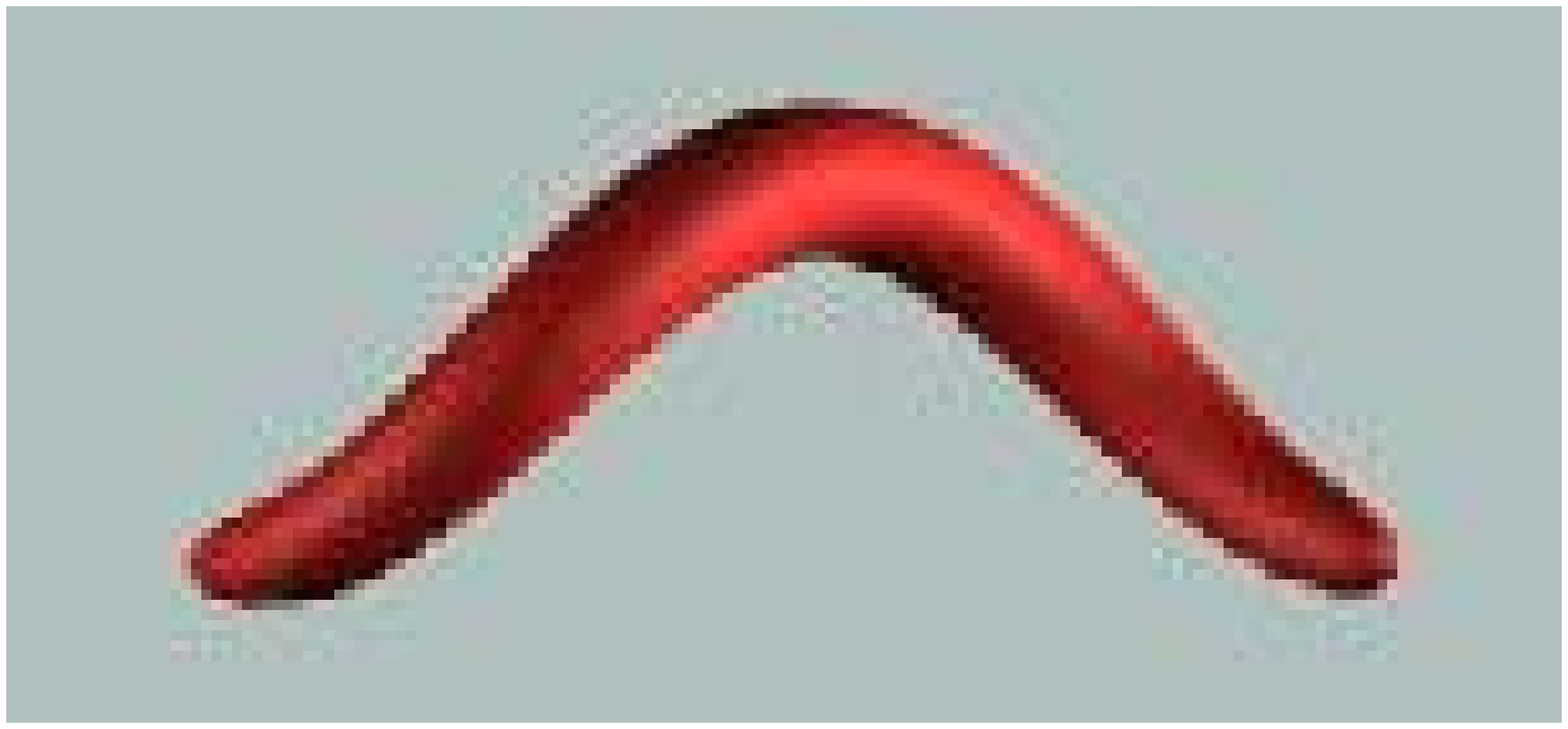,height=1.5cm,width=4.2cm}}
\mbox{\psfig{figure=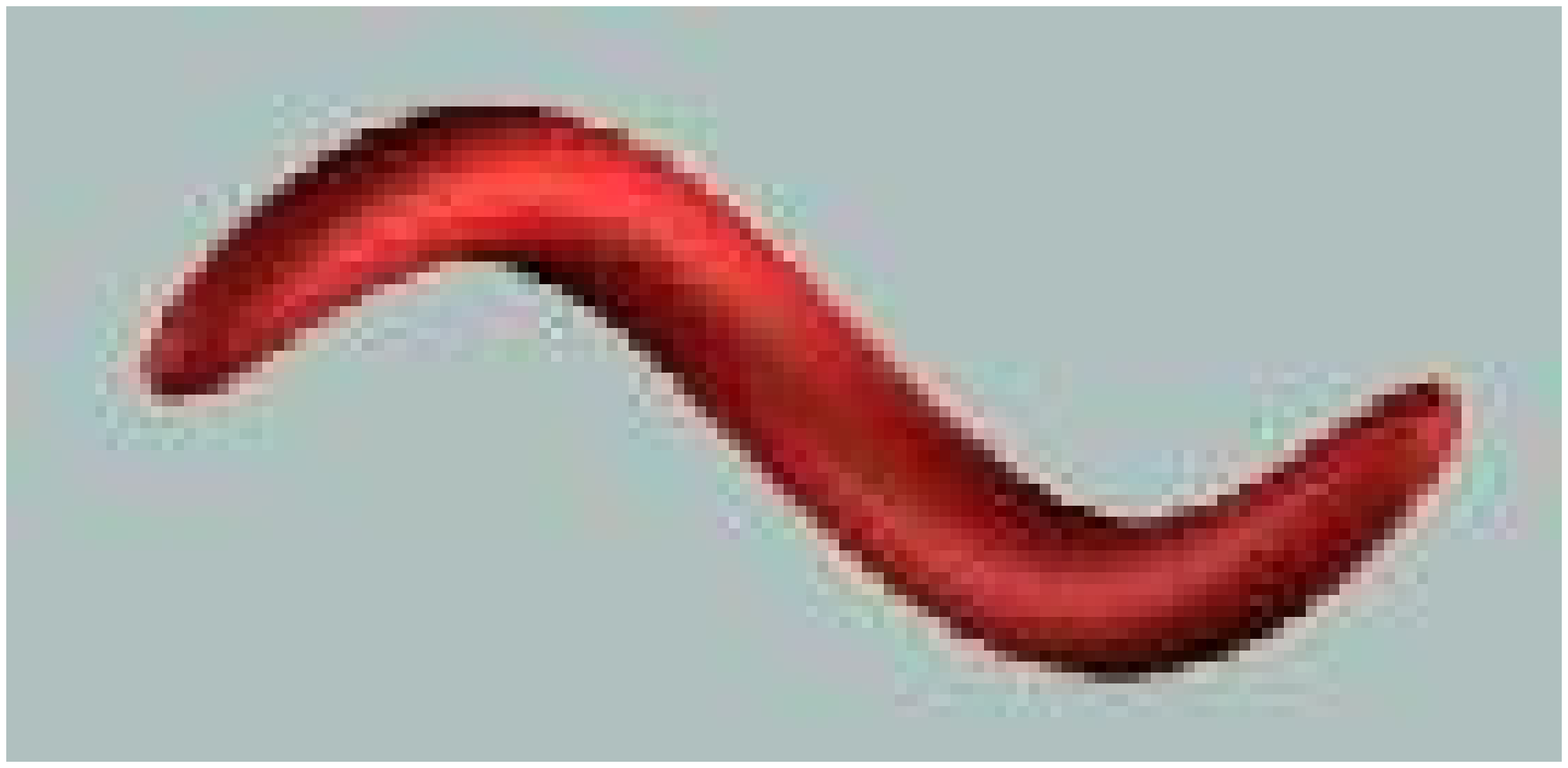,height=1.5cm,width=4.2cm}}}}
\caption{Theoretical example with $\kappa=1$. The translational
propagation is directed left-to-right.}
\end{figure}
\end{center}
 \section{Conclusion}
We introduced and discussed a concept of photon-like object(s) (PhLO) as real,
massless time stable physical object(s) with an intinsically compatible
translational-rotational dynamical structure. So, PhLO are spatially finite ,
with every PhLO a straight-line null direction is necessarily associated, and
the corresponding stress-energy-momentum tensor must be isotropic:
$T_{\mu\nu}T^{\mu\nu}=0$. We showed that in the frame of the relativistic
formulation of vacuum electrodynamics two intrinsically defined strain tensors
$D$ and $D^*$ can be introduced as corresponding Lie derivatives of the flat
Minkowski metric along eigen vectors of $T_{\mu}^{\nu}$. In terms of the
components of these strain tensors can be defined important characteristics of
the internal dynamics of PhLO, in particular, internal energy-momentum exchange
between the $F$-component and $*F$-component of the free field was explicitly
obtained. Definite relations of the projections of $D$ and $D^*$ along the null
direction of translational propagation to the Lie brackets of the electric and
magnetic components were also given. We defined amplitude $\phi$ and phase
$\psi$ of PhLO and showed that the plane wave character of $\phi$ is admissible
and corresponds to the local energy conservation. The NON plane wave character
of $\psi$ guarantees the availability of rotational component of propagation,
which property turned out to be well defined in terms of corresponding
Frobenius curvature. The physical understanding of the Frobenius
nonintegrability of some subcodistributions in this context should read: there
is internal energy-momentum exchange between two individualized subsystems of
PhLO mathematically represented in our case by $F$ and $*F$.

Assuming constant nature of the rotation we come to an extension
of the Riemann-Clifford-Einstein idea for linear relation between
energy-density and Riemannian curvature to linear relation between
energy-density flow and Frobenius curvature as a measure of
nonintegrability.

Finally we note that integrating any of the two 4-forms
$\frac{2\pi l_o}{c}\,\mathbf{d}A\wedge A\wedge \zeta$ and $\frac{2\pi l_o}{c}\,
\mathbf{d}A^*\wedge A^*\wedge \zeta$ over the 4-volume $\mathbb{R}^3\times
l_o$ gives $\pm E.T$, where $T=\nu^{-1}=\frac{2\pi l_o}{c}$ and $E$ is the
integral energy of the solution. This is naturally interpreted as {\it action
for one period}, and represents an analog of the famous Planck formula $E.T=h$.
\vskip 0.5cm
{\bf References}
\vskip 0.2cm
[1]. {\bf M. Born, L. Infeld}, {\it Nature}, {\bf 132}, 970 (1932)

[2]. {\bf M. Born, L.Infeld}, {\it Proc.Roy.Soc.}, {\bf A 144}, 425 (1934)

[3]. {\bf W. Heisenberg, H. Euler}, {\it Zeit.Phys.}, {\bf 98}, 714 (1936)

[4]. {\bf M. Born}, {\it Ann. Inst. Henri Poincare}, {\bf 7}, 155-265 (1937).

[5]. {\bf J. Schwinger}, {\it Phys.Rev}. ,{\bf 82}, 664 (1951).

[6]. {\bf H. Schiff}, {\it Proc.Roy.Soc.} {\bf A 269}, 277 (1962).

[7]. {\bf J. Plebanski}, {\it Lectures on Nonlinear Electrodynamics}, NORDITA,
Copenhagen, 1970.

[8]. {\bf G. Boillat}, {\it Nonlinear Electrodynamics: Lagrangians and
Equations of Motion}, \newline J.Math.Phys. {\bf 11}, 941 (1970).

[9]. {\bf B. Lehnert, S. Roy}, {\it Extended Electromagnetic Theory}, World
Scientific, 1998.

[10]. {\bf D.A. Delphenich}, {\it Nonlinear Electrodynamics and QED},
arXiv:hep-th/0309108, (good review article).

[11]. {\bf B. Lehnert}, {\it A Revised Electromagnetic Theory with Fundamental
Applications}, Swedish Physic Arhive, 2008.

[12]. {\bf D. Funaro}, {\it Electromagnetsm and the Structure of Matter},
Worldscientific, 2008; also: {\it From photons to atoms}, arXiv: gen-ph/1206.3110
(2012).

[13]. {\bf E. Schrodinger}, {\it Contribution to Born's new theory of
electromagnetic feld}, Proc. Roy. Soc. Lond. {\bf A 150}, 465 (1935).

[14]. {\bf G. Gibbons, D. Rasheed}, {\it Electric-magnetic duality rotations in
non-linear electrodynamics}, Nucl. Phys. {\bf B 454} 185 (1995) hep-th/9506035.

[15] {\bf R. Kerner, A.L. Barbosa, D.V. Gal'tsov}, {\it Topics in Born-Infeld
Electrodynamics}, arXiv: hep-th/0108026 v2

[16] {\bf . A. Einstien}
{\it Remarks concerning the essays brought
together in this co-operate volume}, in "Albert Einstein
philosopher-scientist", ed. by P.A.Schillp, The library of living
philosophers, v.{\bf 7}, Evanston, Illinois, 665-688 (1949);

[17] {\bf Stoil Donev, Maria Tashkova}, {\it Geometric View on
Photon-like Objects}, arXiv: math-ph/1210.8323 v1 (now published as a book by
LAP Publishing);

[18]. {\bf J. Marsden, T. Hughes} , (1994), {\it Mathematical
foundations of Elasticity}, Prentice Hall 1983; Reprinted by Dover
Publications, 1994

[19]. {\bf arXiv}:gr-qc/0605025, 0211054, 0605025, 0411145, 0403073, 9701054;
hep-th/9411212 ;math-ph/1312.6751

[20]. {\bf J. Synge}, {\it Relativity: The Special Theory}, North
Holland, Amsterdam, 1958.

\end {document}